\documentclass[
	aps
	,pra
	,twocolumn
	,amsfonts
	,amssymb
	,a4paper
	,showpacs
	]{revtex4}
	
\usepackage{graphicx}

\newcommand{\be}{\begin{equation}}
\newcommand{\ee}{\end{equation}}
\newcommand{\bea}{\begin{eqnarray}}
\newcommand{\eea}{\end{eqnarray}}
\newcommand{\sumlim}{\sum\limits}
\newcommand{\intlim}{\int\limits}

\newcommand{\ket}[1]{|{#1}\rangle}
\newcommand{\bra}[1]{\langle{#1}|}
\newcommand{\proj}[1]{\ket{#1}\bra{#1}}
\newcommand{\braket}[2]{\langle{#1}|{#2}\rangle}
\newcommand{\sandwich}[3]{\langle{#1}|{#2}|{#3}\rangle}

\newcommand{\famvar}{\rho}
\newcommand{\famvartext}{$\famvar$}

\begin{document}


\title{Dynamics of entanglement between two trapped atoms}

\author{Holger Mack}
\author{Matthias Freyberger}

\affiliation{Abteilung f\"ur Quantenphysik, Universit\"at Ulm, D-89069 Ulm, Germany}


\begin{abstract}
We investigate the dynamics of entanglement between two continuous variable quantum systems. The model system consists of two atoms in a harmonic trap which are interacting by a simplified s-wave scattering. We show, that the dynamically created entanglement changes in a steplike manner. Moreover, we introduce local operators which allow us to violate a Bell-CHSH inequality adapted to the continuous variable case. The correlations show nonclassical behavior and almost reach the maximal quantum mechanical value. This is interesting since the states prepared by this interaction are very different from any EPR-like state.
\end{abstract}

\pacs{03.65.-w, 03.65.Ud}

\maketitle


\section{Introduction}

Since the seminal paper \cite{bib:epr} of Albert Einstein, Boris Podolsky and Nathan Rosen (EPR) and the ingenius criticism by John Bell \cite{bib:bell,bib:bellneu} we have begun to understand that quantum correlations can be stronger than ever describable by a classical theory based on local realism. No local realistic theory (LRT) can fully explain all statistical properties predicted by quantum mechanics for specific composite systems. The Hilbert space structure allows for states of quantum systems which we call, according to Erwin Schr\"odinger \cite{bib:schroedinger}, entangled. It was proven that entangled spin-$\frac12$ particles show correlations that violate the so-called Bell inequalities \cite{bib:bellinequalities,bib:wernerwolf}. They provide a boundary on correlations that still can be modelled within a LRT. In fact, for any composite system described by a pure entangled state in a finite Hilbert space a set of observables exists that leads to a violation of the Bell inequalities \cite{bib:violation}. Hence there is a close relation between entanglement and violations of Bell inequalities pointing to the nonclassicality of specific quantum mechanical states.

It became, however, soon clearer that the situation is more complicated when it comes to mixed states described by density operators. Here we need a finer distinction of entanglement. One finds states which are non-separable but still fulfill the Bell inequalities \cite{bib:werner}. Certain classes of these non-separable states can still be subjected to ``entanglement distillation'' and the resulting state then leads to a violation \cite{bib:distillation}. In particular, any non-separable state of two qubits can be distilled \cite{bib:popescubennett}.

But it was also realized \cite{bib:horodecki} that a bipartite system can be in a bound entangled state which satisfies the known versions of Bell inequalities. Therefore we can say that Bell-type arguments can indicate nonclassical correlations of finite-dimensional systems but they cannot characterize all kinds of entanglement \cite{bib:entanglements}.

The situation is again different when we move to continuous variable systems with infinitely large Hilbert space. Correlated continuous variable states can be used as a nonclassical resource for quantum computation and quantum communication \cite{bib:contcomm1,bib:contcomm2,bib:contcomm3}. In particular, it was possible to realize unconditional teleportation in the continuous case \cite{bib:furusawa}. Hence also in this realm entanglement acts as an essential ingredient for nonclassical tasks. Consequently, criteria have been proposed \cite{bib:contentanglement} in order to decide whether a continuous variable state is entangled and protocols have been discussed which purify continuous entangled states \cite{bib:contpurification,bib:parker}.

Composite continuous systems can also violate Bell-type inequalities. One approach is to map the continuous spectrum on a discrete subset of possible measurement outcomes for which the usual form of Bell inequalities applies \cite{bib:banaszek,bib:reid,bib:chen}. This direction has a very appealing phase-space interpretation and it establishes links between nonlocality and the Schr\"odinger-cat gedanken experiment \cite{bib:wodkiewicz}. However, it is also possible to see nonclassical EPR-type correlations in quadrature-phase measurements \cite{bib:reiddrummond} which avoid any discretization.

In the present paper we shall study an elementary continuous variable system: two trapped particles that can collide with each other. In particular we are interested in the dynamical creation of entanglement between the two sub-systems and we shall show that it leads to an almost maximal violation of a Bell-type inequality. Nevertheless, the generated entangled state is completely different from an EPR state. It is actually a superposition of EPR states widely separated in phase space.

The paper is organized as follows: In Sec.~\ref{sec:model} we give a description of the physical model we investigate. The mathematical description in form of a Schr\"odinger equation and its solution is given. In Sec.~\ref{sec:entanglement} we investigate the entanglement contained in the prepared states using an entropic measure. Finally, in Sec.~\ref{sec:chsh} we introduce dichotomic observables which allow us to violate a Bell-type inequality. Sec.~\ref{sec:conclusions} adds the final conclusions.


\section{Model\label{sec:model}}

A variety of quantum mechanical systems can be considered which have continuous degrees of freedom. By this we mean that on such systems we can in principle measure certain observables with a continuous spectrum. Examples are a quantized mode of the electromagnetic field with its quadrature observables and the quantized motion of massive particles with the corresponding position and momentum observables.

Engineering entanglement \cite{bib:engineer} between two such systems and exploiting this entanglement for quantum information purposes requires that the systems can be distinguished via any degree of freedom which does not take part in the entanglement process. If we consider, for example, atoms and entangle their motion, they still can be distinguished via their internal electronic structure.

With current experimental techniques it is possible to trap ultracold atoms in an almost perfect harmonic potential. Moreover, one can address single atoms by exciting their internal structure with the help of lasers. In general the atom-atom interaction is a highly complicated processs. We assume that for cold atoms with a large de-Broglie wavelength we can neglegt the details of the interaction potential and replace it by a pointlike interaction. For a composite system consisting of only two particles with mass $m$ the corresponding Hamiltonian then reads
\be
	\hat H=\frac{1}{2m}\left(\hat p_1^2+\hat p_2^2\right)+\frac12m\omega^2\left(\hat x_1^2+\hat x_2^2\right)+\hbar\gamma\,\delta(\hat x_1-\hat x_2)
	\label{eq:hamiltonian}
\ee
with trap frequency $\omega$ and interaction strength $\hbar\gamma>0$ for a pointlike repulsive potential.

Certain aspects of the dynamics of this model have already been studied \cite{bib:bectwoatoms,bib:twocoldatoms}. Here we shall concentrate on the dynamics of the entanglement induced by the delta interaction. In order to do this we first derive the energies $E$ and eigenfunctions $\ket\Psi$ of the time independent Schr\"odinger equation
\be
	\hat H\ket\Psi=E\ket\Psi
\ee
in appropriate coordinates. If we use a dimensionless center-of-mass coordinate 
\be
  y_c=\kappa_c\cdot\frac12(x_1+x_2)\quad\mbox{with}\quad\kappa_c=\sqrt{\frac{2m\omega}{\hbar}}
  \label{eq:yc}
\ee
and a dimensionless relative coordinate 
\be
  y_r=\kappa_r\cdot(x_1-x_2)\quad\mbox{with}\quad\kappa_r=\sqrt{\frac{m\omega}{2\hbar}}
  \label{eq:yr}
\ee
as well as scaled energies $\tilde E=E/\hbar\omega$ and a dimensionless interaction constant $\tilde\gamma=\gamma\kappa_r/\omega$ we get
\bea
  &&\left[\frac{\partial^2}{\partial y_c^2}+\frac{\partial^2}{\partial y_r^2}-\left(y_c^2+y_r^2\right)\right. \nonumber\\
  &&\qquad\qquad\left.+2\left(\tilde E-\tilde \gamma\,\delta(y_r)\right)\right]\Psi(y_c,y_r)=0.
  \label{eq:scaleequation}
\eea

This Schr\"odinger equation can be solved using a separation ansatz. We present the calculations in Appendix~\ref{app:schroedinger}. We find the eigenvalues $\tilde E_c$ and eigenfunctions $\psi_c(\tilde E_c;y_c)$ for the center-of-mass motion as well as $\tilde E_r$ and $\psi_r(\tilde E_r;y_r)$ for the relative motion of the particles. The total energy then reads $\tilde E=\tilde E_c+\tilde E_r$. We can combine these solutions to get the full solution of the stationary Schr\"odinger equation, Eq.~(\ref{eq:scaleequation}).

The dynamics of the system for scaled times $\tilde t=\omega t$ is then completely given by the wave packet
\bea
	\Psi(y_c,y_r,\tilde t)&=&\sumlim_{\tilde E_c,\tilde E_r}a(\tilde E_c,\tilde E_r)\,\psi_{c}(\tilde E_c;y_c)\,\psi_{r}(\tilde E_r;y_r) \nonumber\\
	&&\qquad\qquad\times\exp\left(-i(\tilde E_c+\tilde E_r)\tilde t\right)
\eea
with complex coefficients $a(\tilde E_c,\tilde E_r)$ determined by the initial condition $\Psi(y_c,y_r,\tilde t=0)$. Rescaling and transforming back the coordinates yields
\bea
	\Psi(x_1,x_2,t) &=& \sqrt{\kappa_r\kappa_c}\cdot\Psi(y_c=\kappa_c\frac{x_1+x_2}2, \nonumber\\
	&&\qquad y_r=\kappa_r(x_1-x_2),\tilde t=\omega t)
	\label{eq:finalsolution}
\eea
for the particle coordinates $x_1$ and $x_2$ at time $t$.

In the following we shall concentrate on the entanglement between the two trapped atoms and in particular on its dynamical evolution due to the pointlike interaction.


\section{Entanglement Measure and Interaction\label{sec:entanglement}}

Most examples of possible applications of entanglement in continuous variable quantum information is based on EPR-states \cite{bib:epr} or in the optical case approximations of EPR-states using squeezed states mixed at a beam splitter \cite{bib:furusawa,bib:silberhorn,bib:ou}. The time evolution in our system now produces entangled states of the two particles quite different from EPR-like states. We will see this by looking at the Wigner function of the relative motion. Nevertheless, within specific time intervals our particles enjoy a strong nonclassical correlation. In order to discuss the strength of the particles' correlation and the nonlocal properties of the corresponding continuous-variable states we have to characterize this entanglement. In this section we will use a measure of entanglement based on the von-Neumann entropy, in Section~\ref{sec:chsh} we present a Bell-test of the nonlocality of the states prepared.

A first hint of the atomic correlation produced by the Hamiltonian $\hat H$, Eq.~(\ref{eq:hamiltonian}), can be given by a common entanglement measure \cite{bib:parker}. For pure states $\ket\Psi$ the von-Neumann entropy 
\be
	S = -\text{tr}\left(\hat\varrho^{\text{red}}\log\hat\varrho^{\text{red}}\right)
	\label{eq:vonNeumannEntropy}
\ee
gives a reasonable measure of entanglement when we insert the reduced density operator
\be
	\hat\varrho^{\text{red}}=\text{tr}\,\proj\Psi.
\ee
Here the trace is taken over one atomic subsystem. Hence in our case we obtain
\be
	\varrho^{\text{red}}(x_1,x_1',t)=\intlim_{-\infty}^{\infty}\!\!dx_2\,\Psi(x_1,x_2,t)\,\Psi^\ast(x_1',x_2,t)
	\label{eq:reduced_space}
\ee
for the state $\Psi(x_1,x_2,t)$, Eq.~(\ref{eq:finalsolution}).

In order to calculate the von-Neumann entropy of $\varrho^{\text{red}}(x_1,x_1',t)$, Eq.~(\ref{eq:vonNeumannEntropy}), we have to diagonalize the reduced density operator in order to find the eigenfunctions $\phi_\nu(x_1,t)$ which fulfill
\be
	\intlim_{-\infty}^{\infty}\!\!dx_1'\,\varrho^{\text{red}}(x_1,x_1',t)\,\phi_\nu(x_1',t)=\lambda_\nu(t)\,\phi_\nu(x_1,t)
	\label{eq:diagonalization}
\ee
with eigenvalues $\lambda_\nu(t)$. Note that these eigenfunctions form a complete and orthonormal set of functions for any time $t$. The integration, Eq.~(\ref{eq:reduced_space}), and the diagonalization, Eq.~(\ref{eq:diagonalization}), can be done numerically using discrete values for $x_1$, $x_2$ and $x_1'$.

With help of the eigenvalues $\lambda_\nu(t)$ and eigenfunctions $\phi_\nu(x_1,t)$ we can write
\be
	\varrho^{\text{red}}(x_1,x_1',t)=\sumlim_{\nu=0}^{\infty}\lambda_\nu(t)\,\phi_\nu(x_1,t)\,\phi_\nu^\ast(x_1',t)
\ee
and the von-Neumann entropy then reads
\be
	S(t) = -\sumlim_{\nu=0}^{\infty}\lambda_\nu(t)\,\log\lambda_\nu(t).
\ee

Our question now is how this entropy of entanglement develops in time starting from an uncorrelated Gaussian product state
\bea
	\Psi(x_1,x_2,t=0)&=&\sqrt{\frac{m\omega}{\pi\hbar}}\exp\left(-\frac{m\omega}{2\hbar}\left(x_1-R/2\right)^2\right) \nonumber\\
	&&\times\exp\left(-\frac{m\omega}{2\hbar}\left(x_2+R/2\right)^2\right)
\eea
for two particles with mean distance $R$. Hence the two atoms start symmetrically in the trap potential with no initial momentum and move towards the trap center where they collide.

When we rewrite this state in the scaled center-of-mass coordinate, Eq.~(\ref{eq:yc}), and the relative coordinate, Eq.~(\ref{eq:yr}), it reads
\bea
	\Psi^{\text{(\famvartext)}}(y_c,y_r,\tilde t=0)&=&\frac1{\sqrt{\pi}}\,\exp\left(-\frac12y_c^2\right)\nonumber\\
	&&\times\exp\left(-\frac12(y_r-\famvar)^2\right).
	\label{eq:initialwavepacket}
\eea
Here we have introduced the scaled distance parameter $\famvar\equiv\kappa_rR$ which now completely characterizes the whole family of dimensionless initial wave packets, Eq.~(\ref{eq:initialwavepacket}). Note further that these initial conditions mean that the center-of-mass motion parametrized by $y_c$ has been cooled to the ground state. Moreover, we recall by looking at Eq.~(\ref{eq:scaleequation}) that the center-of-mass motion and the relative motion decouple. Hence, due to the chosen initial conditions the complete dynamics takes place in the relative motion and our wave function has the product structure
\bea
	\Psi^{\text{(\famvartext)}}(y_c,y_r,\tilde t)&=&\frac1{\pi^{1/4}}\,\exp\left(-\frac12y_c^2\right)\exp\left(-\frac i2\tilde t\right) \nonumber\\
	&&\times\ \psi_r^{\text{(\famvartext)}}(y_r,\tilde t)
	\label{eq:separation}
\eea
for all times $\tilde t$. Consequently, the characteristic features of the system are determined by the wave function $\psi_r$ of the relative motion.

\begin{figure*}
\centerline{\includegraphics{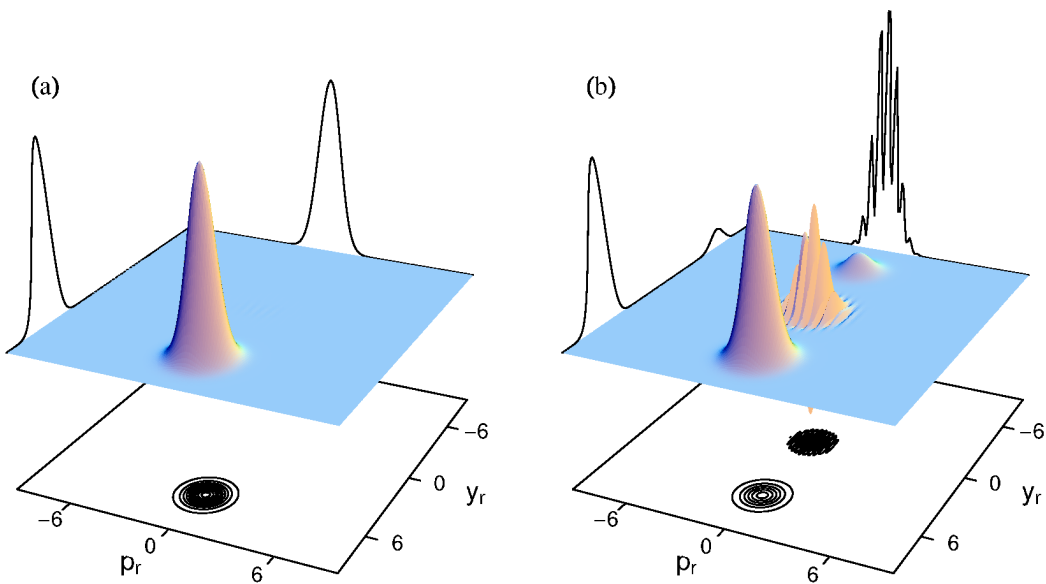}}
\caption{\label{fig:wigner}Wigner function, Eq.~(\ref{eq:wignerfu}), of the relative component $\psi_r^{\text{(\famvartext)}}(y_r,\tilde t)$ of the wave packet, Eq.~(\ref{eq:separation}), at times $\tilde t=0$ (a) and  $\tilde t=4\pi$ (b). The parameters were chosen to be $\famvar=6.0$ and $\tilde\gamma=0.5$.}
\end{figure*}

\begin{figure}
\centerline{\includegraphics{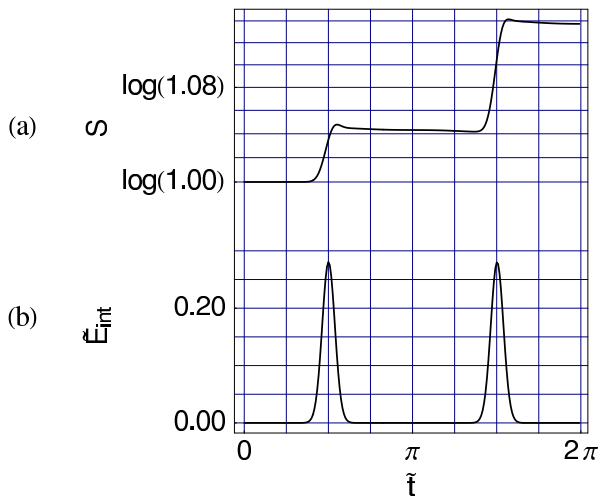}}
\caption{\label{fig:twosteps}Short-time behavior of an initially Gaussian wave packet, Eq.~(\ref{eq:initialwavepacket}), with $\famvar=6.0$ and $\tilde\gamma=0.5$. In the upper graph we plot the entropy $S$, Eq.~(\ref{eq:vonNeumannEntropy}), quantifying the entanglement between the two particles. In the lower graph we show the interaction energy $\tilde E_{\text{int}}$, Eq.~(\ref{eq:einteraction}). The $\tilde t$-axis ranges from $0$ to $2\pi$, the time of one round-trip of two non-interacting particles in a harmonic trap. During the time where the two particles interact, seen explicitely via the increase of interaction energy, the entropy changes, performing a ``step'' towards higher entanglement.}
\end{figure}

First we will have a look at the short-time behavior of this wave packet. Fig.~\ref{fig:wigner} shows the Wigner function
\bea
	W(y_r,p_r,\tilde t)&=&\frac{1}{2\pi}\int\limits_{-\infty}^{\infty}\!\!d\xi\,\exp\left(-ip\xi\right)\,\psi_r^{\text{(\famvartext)}\ast}(y_r-\frac12\xi,\tilde t) \nonumber\\
	&&\times\ \psi_r^{\text{(\famvartext)}}(y_r+\frac12\xi,\tilde t)
	\label{eq:wignerfu}
\eea
of the relative wavefunction $\psi_r$ at times $\tilde t=0$ (a) and $\tilde t=4\pi$ (b). Each time, the wave packet crosses the $\delta$-peak, part of it is reflected coherently while the rest of it passes through the origin. This leads at time $\tilde t=4\pi$ to the humps which are almost Gaussian and opposite to each other in the right part of Fig.~\ref{fig:wigner}. They clearly add up coherently as can be seen from the interference fringes in between.

The state at time $\tilde t=4\pi$ is certainly no longer separable, but it is also much different from an EPR eigenstate. This can be seen if we look for the overlap with a properly scaled EPR-state $\ket{Y_r,P_c}$ derived in App.~\ref{app:overlap}. The main contribution to the overlap, Eq.~(\ref{eq:overlapapp}),
\begin{equation}
	\left|\braket{Y_r,P_c}{\Psi^{\text{(\famvartext)}}(\tilde t)}\right|^2=\left|\psi_r^{\text{(\famvartext)}}(Y_r,\tilde t)\right|^2\cdot\frac1{\sqrt{\pi}}\exp\left(-P_c^2\right)
	\label{eq:overlap}
\end{equation}
is given by the wavefunction $\psi_r^{\text{(\famvartext)}}(Y_r,\tilde t)$ of the relative coordinate. If we now look again at Fig.~\ref{fig:wigner}(b) we can see, that this overlap is not a sharp peak, but obviously delocalized.

Let us understand further how the entanglement builds up in the state Eq.~(\ref{eq:separation}) for short times. In fact we shall see that this is a steplike process: whenever the two particles collide entanglement is added to the system.

In order to see this we show in Fig.~\ref{fig:twosteps}(a) the entanglement $S(\tilde t)$, Eq.~(\ref{eq:vonNeumannEntropy}), as a function of scaled time $\tilde t=\omega t$ together with the average interaction energy
\bea
	\tilde E_{\text{int}}(\tilde t)&=&\frac\gamma\omega\Big\langle\delta(\hat x_1-\hat x_2)\Big\rangle=\tilde\gamma\Big\langle\delta(\hat y_r)\Big\rangle \nonumber\\
	&=&\tilde\gamma\left|\psi_r^{\text{(\famvartext)}}(y_r=0,\tilde t)\right|^2.
	\label{eq:einteraction}
\eea
Here we have used Eqs.~(\ref{eq:finalsolution}) and (\ref{eq:separation}). This energy will help us to identify the time intervals in which the interaction takes place. We have plotted these quantities in Fig.~\ref{fig:twosteps} for time $\tilde t$ ranging from $0$ to $2\pi$. This is exactly the time for a complete round-trip in the case of no interation.

During times of interaction, indicated by a non-vanishing interaction energy $\tilde E_{\text{int}}$, Eq.~(\ref{eq:einteraction}), the entropy changes considerably. As expected, we have an increase in the beginning which goes in a steplike manner. This is characteristic for the short time behavior of the model. Here the main contributions to $\Psi^{\text{(\famvartext)}}$ are still localized and the corresponding Wigner function is not yet smeared out over phase space. We emphasize that for longer times the entanglement can also decrease during the interaction. We discuss this in the next section.

We have seen that the motion of our two particles is dynamically entangled in a steplike process. However, the entanglement measure $S$, Eq.~(\ref{eq:vonNeumannEntropy}), is a rather abstract quantity, which gives us no hint, how this entanglement manifests itself physically. In particular, we shall ask whether this state shows nonlocal behavior under certain circumstances. In the next section we will look for an appropriate situation to answer this question.


\section{Nonclassical Correlations\label{sec:chsh}}


\subsection{Bell-CHSH Inequalities}

In 1964 John Bell gave certain bounds on correlations between two systems within the class of local realistic theories (LRT) \cite{bib:bell,bib:bellneu}. These Bell inequalities were later formulated in different ways, we will use the description by Clauser, Horne, Shimony and Holt (CHSH) \cite{bib:chsh,bib:wernerwolf}.

Bell's inequality is based on local measurements performed on two spatially separated systems $i=1,2$. Within a LRT the crucial assumption is that the observed quantities are described by observables $A_i(\vec\alpha_i,\lambda)$ that only depend locally on the measurement parameters $\vec\alpha_i$ chosen for system $i$ and the complete state $\lambda$ of the total system. As soon as we define $\vec\alpha_i$ and $\lambda$ the value of $A_i$ is determined. If the simplest case is assumed, namely that these values can be only $+1$ or $-1$, the observable is dichotomic. Correlations between the observations are then given by the quantities $\langle A_1(\vec\alpha_1,\lambda)A_2(\vec\alpha_2,\lambda)\rangle_{\text{LRT}}$ obtained by averaging over all possible states $\lambda$ of the corresponding LRT. The Bell-CHSH inequality now limits the range of a combination
\bea
	&&C(\vec\alpha_1,\vec\alpha'_1,\vec\alpha_2,\vec\alpha'_2)\equiv \nonumber\\
	&&\quad\phantom{+}A_1(\vec\alpha_1,\lambda)A_2(\vec\alpha_2,\lambda)
	+A_1(\vec\alpha'_1,\lambda)A_2(\vec\alpha_2,\lambda) \nonumber\\
	&&\quad+A_1(\vec\alpha_1,\lambda)A_2(\vec\alpha'_2,\lambda)
	-A_1(\vec\alpha'_1,\lambda)A_2(\vec\alpha'_2,\lambda)
\eea
of four such correlations to
\be
	|\langle C\rangle_{\text{LRT}}|\leq2
	\label{eq:bellinequality}
\ee
for any choice of the four parameter vectors $\vec\alpha_i$, $\vec\alpha'_i$.

However, if one describes the observed quantities within quantum mechanics, that is by hermitian operators $\hat A_i$ with eigenvalues $\pm 1$ and correspondingly replaces the states $\lambda$ by vectors in Hilbert space or density operators the correlations of the combination
\bea
	&&\hat C(\vec\alpha_1,\vec\alpha'_1,\vec\alpha_2,\vec\alpha'_2)\equiv \nonumber\\
	&&\quad\phantom{+}\hat A_1(\vec\alpha_1)\hat A_2(\vec\alpha_2)
	+\hat A_1(\vec\alpha'_1)\hat A_2(\vec\alpha_2)\nonumber\\
	&&\quad+\hat A_1(\vec\alpha_1)\hat A_2(\vec\alpha'_2)
	-\hat A_1(\vec\alpha'_1)\hat A_2(\vec\alpha'_2),
	\label{eq:cvalueqm}
\eea
are now limited by
\be
	|\langle\hat C\rangle_{\text{QM}}|\leq2\sqrt2.
\ee

The range between $2$ and $2\sqrt2$ lies outside the scope of any LRT and is therefore a clear signature of the nonclassical behavior of the underlying system. In the next paragraph we shall show that within certain time intervals also our system can exhibit such nonclassical features for an appropriate choice of observables.


\subsection{Dichotomic Operators for Wave Packets}

It is not obvious how to transfer the assumptions of the Bell inequality from two-valued discrete systems to continuous systems. Several approaches are conceivable. Their distinction comes from the various kinds of two-valued observables that can be constructed for a continuous system \cite{bib:wodkiewicz,bib:reid,bib:chen}. We will rely on the shifted parity as our measure quantity as proposed in \cite{bib:wodkiewicz}. Note, however, that this is certainly not the only choice.

The parity operator acting on subsystem $i$ reads
\be
	\hat P_i=\intlim_{-\infty}^\infty\!\!dx_i\,\ket{-x_i}\bra{x_i}.
	\label{eq:parity}
\ee
This operator has eigenvalues $\pm1$. A parametrization of this operator is given very naturally by measuring the parity in different reference frames, or equivalently, we use the displaced parity operators \cite{bib:wodkiewicz}
\be
	\hat A_i(x_i,p_i) = \hat D_i^\dagger(x_i,p_i)\,\hat P_i\,\hat D_i(x_i,p_i)
	\label{eq:dispparity}
\ee
for both systems $i=1,2$, that is $\vec\alpha_i=(x_i,p_i)$. The unitary displacement operator
\be
	\hat D_i(x,p) = \exp\left[-\frac i\hbar\left(x\hat p_i-p\hat x_i\right)\right]
\ee
performs a parallel shift mapping the origin $(0,0)$ of phase space to the point $(x,p)$. If we now choose different parameters $(x_1,p_1)$, $(x_1',p_1')$, $(x_2,p_2)$ and $(x_2',p_2')$ we can construct the quantity $\hat C$, Eq.~(\ref{eq:cvalueqm}).


\subsection{Principles of the Measurement}

The physical interpretation of the dichotomic operators $\hat A_1(x_1,p_1)$ and $\hat A_2(x_2,p_2)$, Eq.~(\ref{eq:dispparity}), allows us to describe a procedure to measure these operators for the two particles in the harmonic trapping potential.

The parity is a global property of a quantum state. The operator $\hat P_i$, Eq.~(\ref{eq:parity}), seems to demand a quite complicated measurement process. But if we transform to a representation with energy eigenstates $\ket n_i$ in a symmetric potential for particle $i$ the parity operator reads
\bea
	\hat P_i&=&\intlim_{-\infty}^\infty\!\!dx_i\sumlim_{n=0}^{\infty}\ket{n}_i\braket{n}{-x_i}\bra{x_i} \nonumber\\
		&=&\intlim_{-\infty}^\infty\!\!dx_i\sumlim_{n=0}^{\infty}\ket{n}_i\braket{n}{x_i}(-1)^n\bra{x_i} \nonumber\\
		&=&\sumlim_{n=0}^{\infty}(-1)^n\ket{n}_i\bra{n}.
\eea
From this we see clearly, that the parity operator is dichotomic. A measurement of the parity now means to perform a reduced energy measurement that decides whether $n$ is even or odd. For a state $\ket\psi_i$ of particle $i$ the probability for even parity is given by
\be
	P^{(+)}_i=\sumlim_{n=0}^\infty\left|\,_i\braket{2n}{\psi}_i\right|^2
	\label{eq:pplus}
\ee
and that for odd parity reads
\be
	P^{(-)}_i=\sumlim_{n=0}^\infty\left|\,_i\braket{2n+1}{\psi}_i\right|^2
	\label{eq:pminus}
\ee
which finally results in
\be
	\langle\hat P_i\rangle=P^{(+)}_i-P^{(-)}_i
\ee

The expectation value of the operator $\hat A_i(x_i,p_i)$, Eq.~(\ref{eq:dispparity}), with respect to a state $\ket\psi_i$ can then be determined by measuring the probabilities, Eqs.~(\ref{eq:pplus}) and (\ref{eq:pminus}), for the displaced state $\hat D_i(x_i,p_i)\ket{\psi}_i$. In addition this expectation value $\,_i\sandwich{\psi}{\hat A_i(x_i,p_i)}{\psi}_i$ is up to some scaling the Wigner function $W(x_i,p_i)$ of the state $\ket{\psi}_i$ \cite{bib:wodkiewicz,bib:royer}.

Since we want to measure the parity of both particles simultaneously, we have to separate them in two distinct symmetric traps at measurement time $t$. This prevents any further interaction. Then the trap potentials have to be displaced \cite{bib:trappedions}, and the reduced energy then gives us finally the correlations functions
\be
	\langle\hat A_1(x_1,p_1)\hat A_2(x_2,p_2)\rangle=\sandwich{\Psi}{\hat A_1(x_1,p_1)\hat A_2(x_2,p_2)}{\Psi}
\ee
for the two-particle state $\ket\Psi$. Note here that in the description of this process we have always assumed that the two particles are still distinguishable in a certain degree of freedom. Otherwise it would be impossible to catch them in the appropriate potential.


\subsection{Violation of the Bell-CHSH inequality}

\begin{figure}
\centerline{\includegraphics{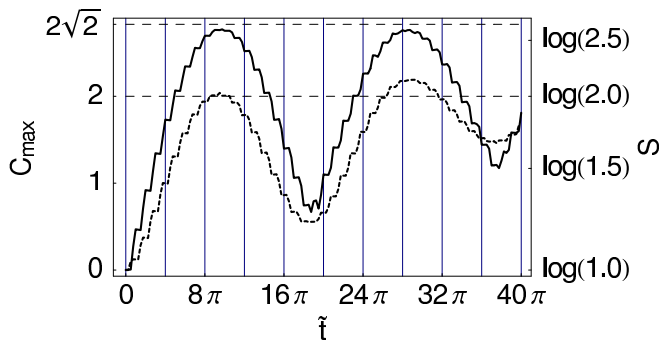}}
\caption{\label{fig:chshplot}Entropy $S$ (dotted line, right axis) and CHSH sum $C_{\text{max}}$ (solid line, left axis) versus time $\tilde t$ for parameters $\famvar=6.0$ and $\tilde\gamma=0.5$. The steps of the entropy, Eq.~(\ref{eq:vonNeumannEntropy}), already seen in Fig.~\ref{fig:twosteps} for short times add up to a oscillating long-time behavior. The same can be seen for the sum $C_{\text{max}}$, Eq.~(\ref{eq:cmax}), of correlations. The bounds for $C_{\text{max}}$ are given by the value $2$ for any local realistic description and $2\sqrt2$ for quantum mechanics. There is clear evidence against a local realistic interpretation of our model within specific time intervals.}
\end{figure}

As stated before, the sum $\hat C$ of correlations, Eq.~(\ref{eq:cvalueqm}), depends on eight real parameters. We used a numerical optimization of these correlations within the full parameter space to get the maximal correlations
\be
	C_{\text{max}}=\max\,\langle\hat C\rangle
	\label{eq:cmax}
\ee
for a given state $\Psi^{\text{(\famvartext)}}$, Eq.~(\ref{eq:separation}). Note that one could also think of varying the initial state, Eq.~(\ref{eq:initialwavepacket}), in order to achieve even higher correlations. We believe, however, that the chosen initial state is the most reasonable one if we assume uncorrelated atoms in the beginning. In Fig.~\ref{fig:chshplot} we show both, the entropic entanglement measure $S$, Eq.~(\ref{eq:vonNeumannEntropy}), and the maximal correlation $C_{\text{max}}$ found for our model, depending on scaled time $\tilde t$.

\begin{figure}
\centerline{\includegraphics{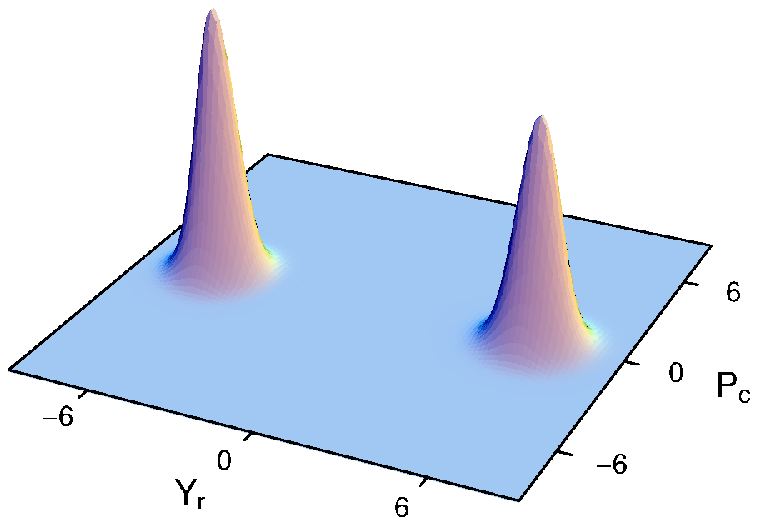}}
\caption{\label{fig:overlap}Overlap, Eq.~(\ref{eq:overlap}), between an EPR-state $\ket{Y_r,P_c}$, Eq.~(\ref{eq:scaledeprstate}), and the state $\ket{\Psi^{\text{(\famvartext)}}(\tilde t=9\pi)}$ with parameters $\famvar=6.0$ and $\tilde\gamma=0.5$. The time is chosen such, that the CHSH-sum $C_{\text{max}}$, Eq.~(\ref{eq:cmax}), takes on its first maximum (see Fig.~\ref{fig:chshplot}).}
\end{figure}

If we look at the entanglement $S$ we still see the single steps discussed before. But now they build up an oscillatory behavior on a much larger time scale and we clearly have time intervals in which atomic collisions decrease the entanglement. The same oscillations also appear in the CHSH correlations. Moreover, they clearly overcome the borders of a local realistic description showing almost maximal violations of the Ineq.~(\ref{eq:bellinequality}).

This strong violation is of particular interest since our dynamically generated state is much different from an EPR-state as already discussed in Sec.~\ref{sec:entanglement}. This becomes clear when we look in Fig.~\ref{fig:overlap} at the overlap, Eq.~(\ref{eq:overlap}), at time $\tilde t=9\pi$, where we find the first maximum of $C_{\text{max}}$ (see Fig.~\ref{fig:chshplot}).


\section{Conclusions\label{sec:conclusions}}

The prospects of quantum information processing using continuous entangled systems clearly demonstrate the need to control the generation of such an entanglement. Presently, not many simple and controllable systems have been described that allow us to start from reasonable initial conditions and to arrive at true nonclassical correlations. We have studied a very elementary model which nevertheless leads to an interesting and dynamically generated entangled state. This state violates a Bell-type inequality when we choose the right set of observables. Moreover, it is much different from the ubiquitous EPR states which have no simple dynamical equivalent and can therefore only be approximated by a physical process with finite resources. In the next step of our study we have to understand how these continuously entangled states behave under reasonable couplings to an environment and we shall devise their potential applications in quantum information processing.

\begin{acknowledgments}
We acknowledge financial support by the Deutsche Forschungsgemeinschaft and by the European Commission through the IST project QUBITS and the IHP network QUEST.
\end{acknowledgments}

\appendix


\section{Solving the Schr\"odinger equation\label{app:schroedinger}}

The time independent Schr\"odinger equation, Eq.~(\ref{eq:scaleequation}), can be divided into two parts. If we use the separation ansatz $\Psi(y_c,y_r)=\psi_c(y_c)\psi_r(y_r)$ the equation splits into two independent ones:
\be
  \left[\frac{\partial^2}{\partial y_c^2}-y_c^2+2\tilde E_c\right]\psi_c(y_c)=0
  \label{eq:comequation}
\ee
and
\be
  \left[\frac{\partial^2}{\partial y_r^2}-y_r^2+2\left(\tilde E_r-\tilde\gamma\,\delta(y_r)\right)\right]\psi_r(y_r)=0
  \label{eq:relequation}
\ee
with constants $\tilde E=\tilde E_c+\tilde E_r$.

Eq.~(\ref{eq:comequation}) for the center-of-mass motion is a harmonic oscillator, not influenced by the particle interaction at all. The well-known eigenvalues $\tilde E_c=n+\frac12$, $n\in\!I\!\!N$, and eigenfunctions
\be
  \psi_{c}(\tilde E_c;y_c)=\left(\sqrt{\pi}2^nn!\right)^{-1/2}H_n(y_c)\,\exp\left(-y_c^2/2\right)
  \label{eq:psicomn}
\ee
form a complete basis for this part of the solution.

Contrarily, Eq.~(\ref{eq:relequation}) contains the interaction part
\be
	\tilde H_{\text{int}}=\tilde\gamma\,\delta(y_r).
	\label{eq:hinteraction}
\ee
The solution therefore splits into two parts --- $\psi_{r,-}(\tilde E_r;y_r)$ for $y_r<0$, $\psi_{r,+}(\tilde E_r;y_r)$ for $y_r>0$ --- of the equation
\be
  \left[\frac{\partial^2}{\partial y_r^2}-y_r^2+2\tilde E_r\right]\psi_{r,\pm}(\tilde E_r;y_r)=0
  \label{eq:relequationnodelta}
\ee
together with two boundary conditions
\be
	\psi_{r,-}(\tilde E_r;y_r=0)=\psi_{r,+}(\tilde E_r;y_r=0)
	\label{eq:boundaries1}
\ee
and
\be
	\psi_{r,+}'(\tilde E_r;y_r=0)-\psi_{r,-}'(\tilde E_r;y_r=0)=2\tilde\gamma\,\psi_{r}(\tilde E_r;y_r=0),
	\label{eq:boundaries2}
\ee
where $\psi'$ denotes the first derivative with respect to $y_r$.

Using the ansatz $\psi_{r,\pm}(\tilde E_r;y_r)=w(z)\,\exp(-z/2)$ with $z=y_r^2$ unveils the differential equation
\be
	z\frac{d^2w(z)}{dz^2}+\left(\frac12-z\right)\frac{dw(z)}{dz}-\frac{1-2\tilde E_r}{4}w(z)=0.
\ee
Solution to this equation is Kummer's function \cite{bib:abramowitz}
\be
	w(z) = U\left(\frac{1-2\tilde E_r}{4},\frac12,z\right)
\ee
which yields
\be
	\psi_{r,\pm}(\tilde E_r;y_r) = {\cal N}_{\pm}U\left(\frac{1-2\tilde E_r}{4},\frac12,y_r^2\right)\exp\left(-y_r^2/2\right)
\ee
with normalization constants ${\cal N}_{\pm}$.

To fulfill the constraints, Eqs.~(\ref{eq:boundaries1}) and (\ref{eq:boundaries2}), we have to look for the derivative
\be
	\psi_{r,\pm}'(\tilde E_r;y_r=0) = \mp 2{\cal N}_\pm\frac{\sqrt\pi}{\Gamma((1-2\tilde E_r)/4)}
\ee
and the value itself
\be
	\psi_{r,\pm}(\tilde E_r;y_r=0) = {\cal N}_\pm\frac{\sqrt\pi}{\Gamma((3-2\tilde E_r)/4)}	
\ee
at the origin of space. Then the conditions can be fulfilled either by
\be
	{\cal N}_+=-{\cal N}_-\quad\mbox{ and }\quad\frac1{\Gamma((3-2\tilde E_r)/4)}=0
	\label{eq:determinodd}
\ee
or
\be
	{\cal N}_+={\cal N}_-\quad\mbox{ and }\quad\frac{2\tilde\gamma}{\Gamma((3-2\tilde E_r)/4)}=-\frac{4}{\Gamma((1-2\tilde E_r)/4)}
	\label{eq:determineven}
\ee
yielding the odd and even eigenfunctions with eigenvalues $\tilde E_r$. While Eq.~(\ref{eq:determinodd}) establishes the undisturbed odd eigenvalues
\be
	\tilde E_r=\frac32,\frac72,\frac{11}2,\dots
	\label{eq:oddev}
\ee
and eigenfunctions
\be
	\psi_{r}(\tilde E_r;y_r)={\cal N}_{\text{odd}}H_{\tilde E_r-1/2}(y_r)\exp\left(-y_r^2/2\right)
\ee
of the harmonic oscillator with normalization
\be
	{\cal N}_{\text{odd}} = \sqrt{\frac{1}{\sqrt{\pi}\,2^{{\tilde E}_r-1/2}\,({\tilde E}_r-1/2)!}},
\ee
the solutions of the transcendental equation
\be
	\quad\frac{\tilde\gamma}{\Gamma((3-2\tilde E_r)/4)}=-\frac{2}{\Gamma((1-2\tilde E_r)/4)}
	\label{eq:transcendental}
\ee
deliver eigenvalues and even eigenfunctions
\bea
	\psi_{r}(\tilde E_r;y_r)&=&{\cal N}_{\text{even}}\ U\left(\frac{1-2\tilde E_r}{4},\frac12,y_r^2\right) \nonumber\\
	&&\times\ \exp\left(-y_r^2/2\right)
	\label{eq:eveneigenfunction}
\eea
with
\be
	{\cal N}_{\text{even}} = \sqrt{\frac{\Gamma\left(\frac34-\frac{{\tilde E}_r}2\right)\Gamma\left(\frac14-\frac{{\tilde E}_r}2\right)}{\pi\left[\psi\left(\frac34-\frac{{\tilde E}_r}2\right)-\psi\left(\frac14-\frac{{\tilde E}_r}2\right)\right]}},
\ee
where $\psi(x)$ is the logarithmic derivative of $\Gamma(x)$

\begin{figure}
\centerline{\includegraphics{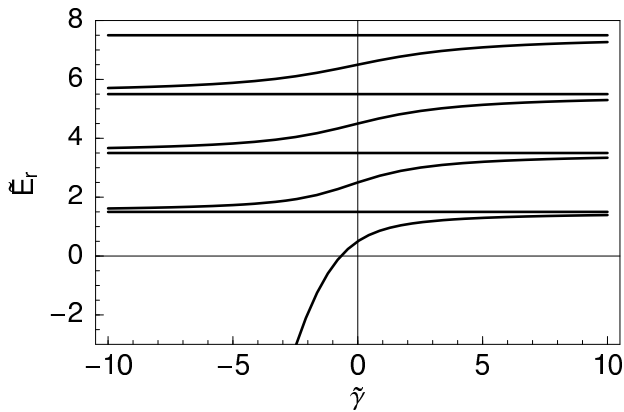}}
\caption{\label{fig:evals}Eigenvalues $\tilde E_r$, Eq.~(\ref{eq:oddev}) and solutions of Eq.~(\ref{eq:transcendental}), for the Schr\"odinger equation, Eq.~(\ref{eq:relequation}), of the relative motion of the two particles with $\delta$-like interation. The interaction parameter $\tilde\gamma$ can be either positive, meaning repulsive interation, or negative, corresponding to an attractive force. While the odd numbered eigenvalues remain constant with varying $\tilde\gamma$, the other eigenvalues increase with stronger repulsion.}
\end{figure}

In Fig.~\ref{fig:evals} we show the dependence of the eigenvalues of these solutions on the interation strength $\tilde\gamma$. For no interaction ($\tilde\gamma=0$) the values correspond to the equally spaced eigenvalues of the harmonic oscillator. The energy eigenvalues increase due to repulsion or decrease for attractive interaction.

\begin{figure}
\centerline{\includegraphics{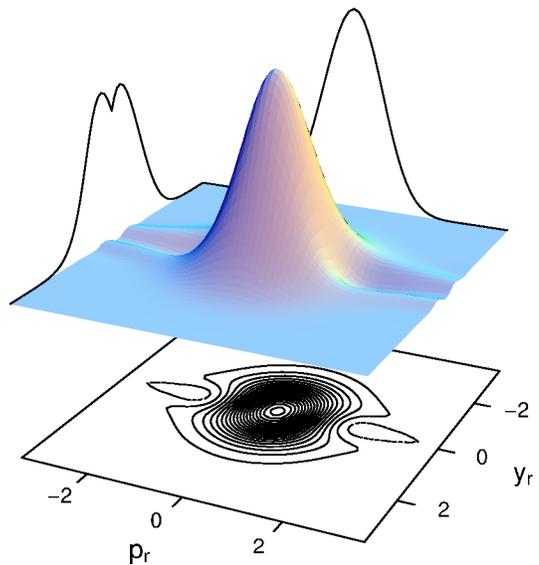}}
\caption{\label{fig:groundstate}Wigner function of the ground state of the radial wave function $\psi_{r}(\tilde E_r;y_r)$, Eq.~(\ref{eq:eveneigenfunction}), for an interaction of the form $\tilde H_{\text{int}}=\tilde\gamma\delta(y_r)$ with parameter $\tilde\gamma=0.5$. The energy of this ground state, determined by Eq.~(\ref{eq:transcendental}), is given by $\tilde E_r\approx0.7335$, well above the value $0.5$ for the non-interacting case. We find a perturbed Gaussian distribution having negative values along the axis $y_r=0$.}
\end{figure}

The Wigner function of the ground state is shown in Fig.~\ref{fig:groundstate}. One can see a clear dent in the position distribution located at the origin, which indicates the repulsive interaction.


\section{EPR-States in position representation\label{app:overlap}}

The EPR-State \cite{bib:epr} for two subsystems $1$ and $2$
\bea
	\ket{X,P}&=&\frac1{\sqrt{2\pi\hbar}}\int\!\!dx_1\int\!\!dx_2\ \exp\left(iPx_1/\hbar\right) \nonumber\\
	&&\quad\times\ \delta(x_1-x_2-X)\ket{x_1}\ket{x_2}
\eea
fulfills the eigenvalue equations
\be
	\left[\hat x_1-\hat x_2\right]\ket{X,P}=X\ket{X,P}
\ee
and
\be
	\left[\hat p_1+\hat p_2\right]\ket{X,P}=P\ket{X,P}.
\ee
If we use Eqs.~(\ref{eq:yc}) and (\ref{eq:yr}) we can define scaled EPR-states
\bea
	\ket{Y_r,P_c}&=&\frac1{\sqrt{2\pi}}\int\!\!dy_r\int\!\!dy_c\ \exp\left(iP_cy_c\right) \nonumber\\
	&&\quad\times\ \delta(y_r-Y_r)\ket{y_r}\ket{y_c}
	\label{eq:scaledeprstate}
\eea
which are now given in the relative coordinate $y_r$ and the center-of-mass coordinate $y_c$ with parameters $Y_r\equiv\kappa_rX$ and $P_c\equiv\frac{\kappa_c}{2\hbar}P$.

We can now compare the state of our two atoms to such an EPR-state. We take the state $\Psi^{\text{(\famvartext)}}(y_c,y_r,\tilde t)$ from Eq.~(\ref{eq:separation}), and find the product
\bea
	\braket{Y_r,P_c}{\Psi^{\text{(\famvartext)}}(\tilde t)}&=&\psi_r(Y_r,\tilde t)\cdot\frac1{\sqrt{2}\pi^{3/4}}\!\int\!\!dy_c\exp\left(-iP_cy_c\right) \nonumber\\
	&&\times\ \exp\left(-\frac12y_c^2\right)\exp\left(-\frac i2\tilde t\right)\nonumber\\&=&\psi_r(Y_r,\tilde t)\cdot\frac1{\pi^{1/4}}\exp\left(-\frac12P_c^2\right) \nonumber\\
	&&\times\ \exp\left(-\frac i2\tilde t\right).
\eea
This leads to the overlap
\be
	\left|\braket{Y_r,P_c}{\Psi^{\text{(\famvartext)}}(\tilde t)}\right|^2=\left|\psi_r(Y_r,\tilde t)\right|^2\cdot\frac1{\sqrt{\pi}}\exp\left(-P_c^2\right)
	\label{eq:overlapapp}
\ee
which is a Gaussian in the variable $P_c$ and reflects the probability distribution of the wave function of the relative coordinate.


\end{document}